Zukic Dž., Egger J., Bauer M. H. A., Kuhnt D., Carl B., Freisleben B., Kolb A., Nimsky Ch.

# Glioblastoma Multiforme Segmentation in MRI Data with a Balloon Inflation Approach

**Abstract**

*Gliomas are the most common primary brain tumors, evolving from the cerebral supportive cells. For clinical follow-up, the evaluation of the preoperative tumor volume is essential. Volumetric assessment of tumor volume with manual segmentation of its outlines is a time-consuming process that can be overcome with the help of computer-assisted segmentation methods. In this paper, a semi-automatic approach for World Health Organization (WHO) grade IV glioma segmentation is introduced that uses balloon inflation forces, and relies on the detection of high-intensity tumor boundaries that are coupled by using contrast agent gadolinium. The presented method is evaluated on 27 magnetic resonance imaging (MRI) data sets and the ground truth data of the tumor boundaries – for evaluation of the results – are manually extracted by neurosurgeons.*

**Introduction**

Gliomas are the most common primary brain tumors, whereof 70% are among the group of malignant gliomas (anaplastic astrocytoma World Health Organization (WHO) grade III, glioblastoma multiforme WHO grade IV) [11]. The glioblastoma multiforme (WHO IV) is one of the highest malignant human neoplasms. Due to the biological behavior, gliomas of WHO grade II to IV cannot be cured with surgery alone. The multimodal therapeutical concept involves maximum safe resection followed by radiation and chemotherapy, depending on the patient's Karnofsky[1] scale. The survival rate still only accounts approximately 15 months [12], despite new technical and medical accomplishments such as multimodal navigation during microsurgery, stereotactic radiation or the implementation of alkylating substances. Although there is still a lack of Class I evidence, literature today favors a maximum extent of resection in low- and high-grade gliomas as a positive predictor for longer patient survival [13].

The clinical follow-up of the tumor volume is essential for an adaptation of the therapeutical concept. Therefore, the exact evaluation is fundamental to reveal a recurrent tumor or tumor-progress as early as possible. Volumetric assessment of a tumor with manual segmentation of its outlines is a time-consuming process that can be overcome with the help of computer-assisted segmentation methods.

In the following section, related work for tumor or especially glioma segmentation based on magnetic resonance imaging (MRI) data sets is summarized. For a comprehensive overview of some deterministic and statistical approaches see the review of Angelini et al. [1].

Gibbs et al. [8] presented a combination of region growing and morphological edge detection for segmentation of contrast enhancing tumors in T1 weighted MRI data sets. Starting with a manually provided first sample of tumor signal and surrounding tissue, an initial segmentation using pixel thresholding, morphological opening and closing and fitting to an edge map is performed. Gibbs et al. evaluated their procedure with one phantom data set and ten clinical data sets. However, the mean segmentation time for a tumor was about ten minutes. They did not exactly classify the tumors they used for their evaluation.

An interactive method for segmenting full-enhancing, ring-enhancing and non-enhancing tumors has been presented by Letteboer et al. [14]. They evaluated their approach with twenty clinical cases. Based on a manual tracing of an initial slice, morphological filter operations are applied to the MRI volume to divide the data in homogenous regions.

A deformable model depending on intensity-based pixel probabilities for tumoral tissue has been introduced by Droske et al. [5]. They used a level set formulation, in order to split the MRI data into regions of similar image properties for tumor segmentation. The method was then performed on image data of twelve patients.

Clark et al. [3] proposed a knowledge-based automated segmentation on multispectral data to partition glioblastomas. After a training phase with fuzzy C-means classification, clustering analysis and a brain mask computation, initial tumor segmentation from vectorial histogram thresholding is post processed to eliminate non-tumor pixels. The introduced system has been trained on three volume data sets and has been tested on thirteen unseen volume data sets.

A segmentation based on outlier detection in T2 weighted MRI data has been developed by Prastawa et al. [15]. Therefore, in order to detect abnormal tumor regions, the image data is registered on a normal brain atlas. Then, tumor and edema are isolated by statistical clustering of the differing voxels and a deformable model. However, they have applied the method to three real data sets. For each case, the required time for automatic segmentation was about 90 minutes.

Sieg et al. [17] proposed an approach for segmenting contrast-enhanced, intracranial tumors and anatomical structures of registered multispectral MRI data. Multilayer feed-forward neural networks with back-propagation are trained and a pixel-oriented classification is applied. The approach has been tested on twenty-two data sets, but the authors did not provide any computational time.

Egger et al. [6,7] introduced a segmentation scheme for spherical objects that creates a directed 3D graph by sending rays through the surface points of a polyhedron

---
[1] Scale of 0-100, indicating functional impairments of the patient (0=dead, 100=healthy)

and sampling the graph's nodes along each ray. Thereafter, the minimal cost closed set on the graph is computed via a polynomial time s-t cut [2], creating an optimal segmentation of the tumor. The center of the polyhedron is defined by the user and located inside the tumor (e.g. glioma).

The paper is organized as follows. Section 2 presents the details of the proposed approach. In Section 3, experimental results are presented. Section 4 discusses the paper and outlines areas for future work.

**Material and Methods**

The overall method relies on user initialization. The user draws an approximate outline on a slice that is approximately located central to the tumor (Figure 1, left side). From this initialization, the following information is extracted:

1. Center of the tumor: Two coordinates (x,y) are extracted from the center of the outlined object (center of area), and the third coordinate (z) is the index of the selected slice.

2. By analyzing pixel intensities in the selected slice, the minimum and maximum intensities of voxels of interest are determined, ignoring few highest and lowest percent in order to account for noise.

3. The average distance from the center to the boundary: The average distance from the center to the boundary is a scale-invariant measure – it is the same in 2D (slice) and 3D (whole volume), e.g. the radius of a circle is equal to the radius of the sphere made by rotating that circle around its diameter.

The main idea is to start with a small triangular surface mesh in the shape of a convex polyhedron at the approximate center of the glioma. Balloon inflation forces [4] are used to expand this mesh, keeping it approximately star-shaped[2]. We do not inflate beyond the glioma boundary, and declare that the segmentation is finished when the increase in average center-surface distance slows down. The following steps are executed iteratively:

1. Split polyhedron edges that are 3 times longer than average voxel spacing (geometrical mean of spacing in X, Y and Z direction).

2. Compute per-vertex surface normals and curvature estimates.

3. Do inflation, i.e. for each vertex:

    a) Calculate the cosine of the angle ($\varphi$) between center-vertex vector ($\vec{d}_{cv}$) and surface normal vector ($\vec{n}_v$) at the given vertex - the greater the angle, the lower the inflation speed. While the surface is still nearly spherical, i.e. at the start of inflation process, the inflation speed is higher and thus the surface reaches the boundary area faster.

    b) Calculate the move-speed factor. The higher the curvature, the lower the inflation speed, thus the inflation speed is slowed down for vertices on ridges, valleys and peaks.

    c) If a vertex can be moved, it is moved in the direction of center-vertex vector (thus maintaining star-shape). Displacement amount is adjusted by inflation speed factors (Figure 1 right).

4. Smooth the surface of the polyhedron slightly. This is required to overcome noisy voxels, which would otherwise prevent inflation of the mesh beyond them, even if they are in the middle of the glioma. We also know that the surface of the glioma is smooth (and not jaggy).

The vertex can be moved if the target position (current position + displacement outwards) has an intensity within the initialized range of interest (range of values within user initialized boundary), and the target intensity is higher than 80% of the maximum intensity this vertex has encountered so far. This additional condition prevents inflation outside of the boundary region, which has higher intensity than normal brain tissue and tissue inside the glioma. This is a consequence of the use of a gadolinium contrast agent.

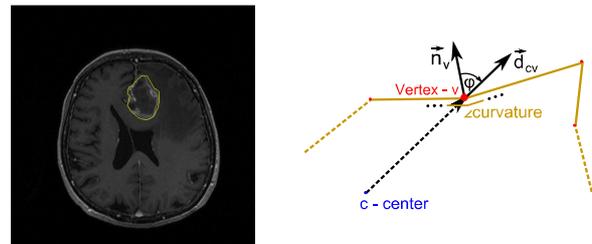

Fig. 1: Left: User initialization of a tumor boundary (yellow contour). Right: Features involved in vertex movement calculation ($\vec{n}_v$ - normal at vertex v, $\vec{d}_{cv}$ - center-vertex vector).

**Results**

The presented methods were implemented in C++. The segmentation in our implementation took about 1 second per data set on an Intel Core i7-920 CPU, 2.66, on Windows7 x64. To evaluate the approach, neurological surgeons with several years of experience in the resection of tumors performed manual slice-by-slice segmentation of 27 WHO grade IV gliomas. The tumor outlines for the segmentation were displayed by the contrast-enhancing areas in T1 weighted MR images. Afterwards, the segmentation results were compared with the segmentation results of the proposed method via the Dice Similarity Coefficient (DSC) [16,18]. The average DSC for all data sets was 80.46% (see Table 1). Figure 2 shows the segmentation results of the presented approach (from the upper left to the lower right): two axial slices, one sagittal slice and one 3D visualization. Figure 3 presents a direct comparison of two manual and automatically segmented MRI slices. The images on the left show the original slices with the glioblastoma multiforme. The two slices in the middle show the manual segmentation and the two slices on the right show the results of the automatic segmentation with the balloon inflation approach.

---

[2] A star-shaped object is an object in which a point (the center) exists that can be connected with every surface point by a straight-line segment, and all points of those straight-line segments lie entirely within the object. A star is a representative 2D example of this class of objects. All convex objects are also star-shaped objects (but not vice versa).

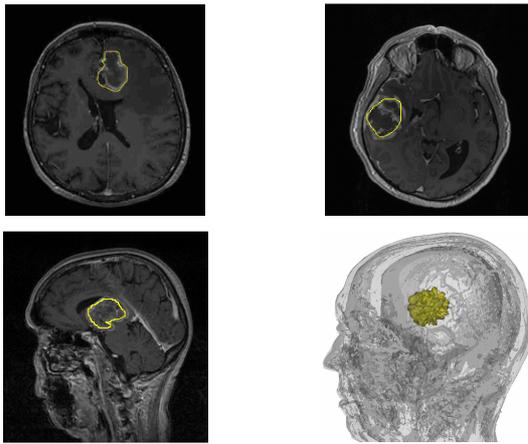

Fig. 2: Results of automatic tumor segmentations: The lower left image belongs to the 3D model of the tumor (lower right image).

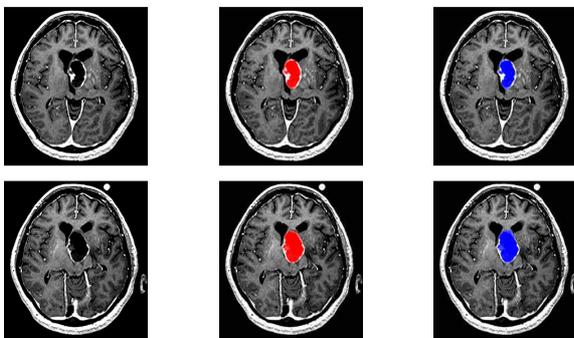

Fig. 3: Comparison of two slices: native (left), manual segmentation (middle) and automatic segmentation (right).

Table 2 and Figure 4 present the segmentation results for different user initializations for a selected data set. The values belong to a tumor that was located in a MRI data set between slice number 42 and slice number 73 (center slice for the tumor: 57/58). As you can see the DSC decreases if the user draws the initial contour on a slice that is located near the border of the tumor. However, if the user selects one of the slices around the tumor center (55-60), the resulting DSC is over 80%. So there is a certain robustness of the presented approach if the user doesn't select a slice that is located too far away from the tumor center. We even varied the user initialization of one slice (no. 64) and got the following DSC results: 82.03%, 85.76%, 80.65%, 84.68% and 82.47%.

TABLE II
SEGMENTATION RESULTS FOR DIFFERENT USER INITIALIZATIONS.

| slice | Tumor volume (mm$^3$) | | Number of voxels | | DSC (%) |
|---|---|---|---|---|---|
| | manual | algorithm | manual | algorithm | |
| 45 | | 2532.96 | | 21758 | 21.7 |
| 50 | | 9568.28 | | 82191 | 62.7 |
| 55 | | 15636.4 | | 134316 | 82.81 |
| 56 | | 16625.9 | | 142816 | 82.96 |
| 57 | | 15768.7 | | 135452 | 84.33 |
| 58 | | 18457.6 | | 158550 | 84.9 |
| 59 | 16259.7 | 19005.9 | 139670 | 163260 | 85.39 |
| 60 | | 18441.8 | | 158414 | 86.74 |
| 61 | | 20612.1 | | 177057 | 82.66 |
| 62 | | 21294.2 | | 182916 | 82.4 |
| 63 | | 18321.6 | | 157382 | 76.77 |
| 64 | | 18236.4 | | 156650 | 80.66 |
| 65 | | 20758.3 | | 178313 | 76.71 |
| 70 | | 6714.59 | | 57678 | 35.38 |

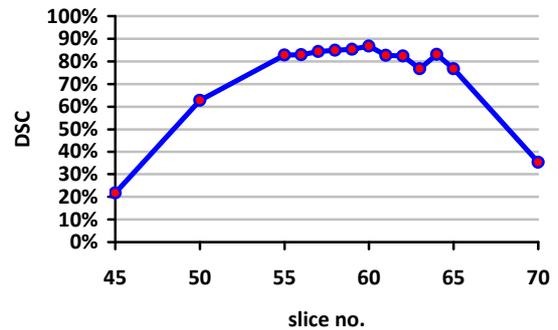

Fig. 4: DSC results for different user initializations. The lower tumor boundary for this data set was slice number 42 and the upper tumor boundary was slice number 73.

**Discussion**

In this paper, a semi-automatic approach for WHO grade IV gliomas (glioblastoma multiforme) has been presented. The introduced method uses a new segmentation scheme that is based on balloon inflation forces. The presented approach has been evaluated on 27 magnetic resonance imaging data sets with WHO grade IV gliomas. Experts (neurosurgeons) with several years of experience in resection of gliomas extracted the tumor boundaries manually to obtain the ground truth of the data. The manually segmented results and the results of the semi-automatic approach have been compared with each other by calculating the average Dice Similarity Coefficient.

TABLE I
SUMMARY OF RESULTS: MIN., MAX., MEAN AND STANDARD DEVIATION FOR 27 GLIOMAS.

| | Volume of tumor (cm$^3$) | | Number of voxels | | DSC (%) | manual segmentation time (min.) |
|---|---|---|---|---|---|---|
| | manual | algorithm | manual | algorithm | | |
| min | 0.79 | 0.77 | 1526 | 2465 | 63.72 | 3 |
| max | 73.45 | 73.32 | 550307 | 446560 | 94.02 | 19 |
| $\mu \pm \sigma$ | 21.64 $\pm$ 19.16 | 20.25 $\pm$ 19.27 | 100349.93 | 86589.70 | 80.46 $\pm$ 7.42 | 6.93 $\pm$ 4.11 |

For an accurate tumor volume it is necessary to develop methods – like the one introduced in this contribution – that use all slices to calculate the tumor boundaries. Simpler methods like geometric models provide only a rough approximation of the tumor volume and should not be used, when accurate determination of size is of paramount importance in order to draw safe conclusions in oncology. Geometric models use one or several user-defined diameters – which can be manually achieved very quickly – to calculate the tumor volume. Briefly, according to the spherical model, the volume is defined as $1/6\ \pi d^3$ (d is the diameter of the maximum cross-sectional area) and the ellipsoid model defines the volume as $1/6\ \pi abc$ (a, b, c represent the diameters in the three axes of the tumor) [10].

There are several areas of future work. For example, the presented segmentation scheme can be enhanced with statistical information about shape and texture of the desired object [9]. Furthermore, the method can be evaluated on MRI data sets with WHO grade I, II and III gliomas. Additionally, the approach will be compared with a recently introduced graph based approach [6].

## Acknowledgements

The authors would like to thank Fraunhofer MeVis in Bremen, Germany, for their collaboration and especially Horst K. Hahn for his support.

## Affiliation of the first Author

Dženan Zukić
University of Siegen, Computer Graphics Group
Hölderlinstraße 3, 57076 Siegen, Germany
Phone +49 271 740 2826
zukic@fb12.uni-siegen.de